# Automatic, Real-time Classification of User Feedback Using Large Language Models

JIM MADDOCK, ROSE LEITNER, and ANNA WU

In this paper we discuss an ongoing multi-year project that aims to make open text feedback more accessible and useful to UX practitioners by automating classification and providing real time access to comments, themes, and analysis. By significantly lowering the time and knowledge cost of implementing automated solutions, we aim to effectively democratize our data analysis processes, allowing and encouraging non-technical stakeholders to access and leverage data on their own. We share both the organizational and technical constraints we have encountered over the course of this project, and the solutions we have prototyped as a result of those constraints.

## 1 Introduction

Understanding user sentiment is an important part of the product development cycle. While UX researchers typically engage in foundational and evaluative research before a product launch in order to understand *what* to build and *how* users will interact with it, monitoring and evaluating user sentiment after a launch is equally important for continued improvement and development. Through continuous evaluation, researchers can (and should) *update* UX research Points of View (PoV)[1] established during the initial development cycle[4].

As UX Researchers working at Google, we often leverage various channels (e.g. ad-hoc surveys, in-product surveys, and supportability cases) to collect user feedback continuously after a product launch. Open text responses constitute a particularly rich and potentially insightful form of user feedback, but these data are often underutilized or even ignored due to the high time cost required for evaluation.

In this workshop paper we discuss an ongoing multi-year project that aims to make open text feedback more accessible and useful to researchers and product teams by automating classification and providing real time access to comments, themes, and analysis. By significantly reducing the time and knowledge cost of implementing automated solutions, we aim to effectively democratize our data analysis processes, allowing and encouraging non-technical

[1]Dogan et al. define a UX research PoV as "a perspective based on data, evidence and insights," which UX researchers must develop in order "to clearly represent the core user requirements while balancing this with an appreciation of technical constraints and business needs." For a detailed explanation of the PoV framework and playbook which we reference throughout this paper, see the original workshop proposal.

Authors' Contact Information: Jim Maddock, maddock@google.com; Rose Leitner, mleitner@google.com; Anna Wu, annwu@google.com.

stakeholders to access and leverage data on their own. In the language of Dogan et al.[4], we contribute a "play" to the UX research playbook by sharing both the organizational and technical constraints we have encountered over the course of this project, and by outlining the solutions we have prototyped as a result of those constraints.

This project has resulted in several different tools designed for different purposes, which we discuss in the main body of the paper. Broadly speaking, these systems *classify* open text data into categories. Additionally, some systems *extract themes* before classifying each open text response. For simplicity, we focus exclusively on classification systems. As we further develop these tools, we continue to improve automation and reduce the technical knowledge required by end users to operate and maintain these systems.

Our goal in writing this workshop paper is to describe our practical experience as developers of low barrier to entry open text classification systems, such that other industry practitioners can learn from our process. While prior academic work documents the performance trade-offs of using these systems[1–3, 5], we focus on the process of implementing classification pipelines at scale across multiple teams in an industry setting. Ultimately we view these systems as a powerful "play" in the UX research playbook, which will allow industry practitioners to leverage underutilized data, to generate insights, and ultimately to better understand and describe their users.

## 2 In-product Surveys for Collecting User Feedback

Data collection is one of four key building blocks necessary to establish a UX research PoV, situated between foundation and insight generation [4]. In industry, we often leverage in-product surveys to collect user feedback continuously after a product launch. In-product surveys can take many different forms, but they typically contain some combination of multiple choice questions ("Overall, how satisfied are you with <product>?"), as well as open text fields ("What, if anything, do you most dislike about <product>?"). Users see the survey as a small pop-up in the bottom corner of their browser, which they can choose to interact with or ignore.

While multiple choice and Likert scale responses from in-product surveys are frequently leveraged to inform business leadership and drive product decisions, these types of measures have a number of drawbacks. In our experience, purely quantitative measures collected outside of a/b experiments are difficult to interpret at any given point in time, and they are difficult to change from one collection period to another. For instance, if the mean satisfaction score is 4 on a scale of 1 to 5 during a specific collection period, it's challenging to derive much practical meaning from that score. Furthermore, these quantitative measures provide little actionable insight when it comes to developing and improving products. We may know that our users are "moderately satisfied" with a given product, but this metric provides little guidance towards what our engineers should build.

Open text responses often provide deeper insight than a simple multiple choice response, but typically require far more effort to analyze. In response to the question "What, if anything, do you most dislike about <product>?" users can not only tell us whether they are satisfied with a product, but also why they are unsatisfied. Collecting open text feedback through in-product surveys provides some of the fidelity of an interview while also providing the scalability and continuous data collection of a survey. This fidelity leads to potentially deeper and more meaningful synthesis during insights generation—the building block following data collection—but the effort required to move between data collection and insights generation is also far higher.

Despite its potential to surface deeper insights, open-text feedback is under-utilized by our research teams due to high time cost. At a minimum, researchers must read and categorize each individual response in order to develop an accurate summary of open text user feedback. Furthermore, many open text responses are off topic or unhelpful and must be manually filtered and discarded. In our product area alone, in-product surveys provide hundreds to thousands of responses within a single month, and within our fast paced development cycle which is both time and resource constrained, researchers often cannot take advantage of these types of feedback. Increases in the number of end users and the appetite for customer data continue to exacerbate this problem, making manual analysis of open text data less and less feasible as products continue to scale. As a result, these rich data are never leveraged to develop a PoV, and the process stalls at the data collection phase.

## 3 Automating Open-text Classification with Machine Learning

In response to these challenges, we kicked off a small machine learning based automation project late in 2022 that aimed to automatically classify and extract sentiment from open text responses. Our initial goal was to simply reduce the time and effort needed to manually code text. As we've continued to productionize, automate, and expand the tools we develop, we've also attempted to reduce technical barriers so that researchers and stakeholders can leverage open text feedback as a data source for a range of applications, including reports, dashboards, and real-time monitoring tools.

As a first approach towards automating open text feedback classification, we leveraged a series of traditional machine learning classifiers built into a data processing pipeline. Our system automatically filtered spam or off topic responses, and it categorized helpful responses into one of several pre-defined categories.

Initially, we aimed to reduce the time and effort needed to manually code text simply by eliminating the need for a human to manually code each item. As we interviewed researchers and other UX practitioners across Google, however, we also realized we needed to prioritize flexibility, extensibility, and scalability in implementation. We found that a few other teams had already developed solutions, but that these other systems were either expensive, or they were difficult to extend to other product areas. In other words, even if we reduced the time cost associated with manually coding feedback, trading manual labor for a high technical implementation cost would result in low uptake.

We therefore designed a low-cost system that could be adapted and retrained for other product areas, and that utilized different categorization schemes without extensive engineering knowledge or time investment. The system architecture was based on a modular design that could expand to different classes and products. Each category or theme was predicted by a single binary classifier, and a classification pipeline was made up of a collection of classifiers. Each product area would then maintain its own pipeline, which would allow products to select which classifiers to include in their prediction pipeline.

In order to reduce the startup cost of implementing this system, we used an internal auto-ML platform based on TensorFlow[2] for training and hosting models. Once trained, these models could be easily accessed by our internal data pipelining system, allowing near real-time classification of incoming data. We envisioned (and built) both a library of existing classifiers, and simple scripts to train new classifiers. Users could start with existing classifiers that work across products. For categories that are product specific, users could provide training data and train their own product classifiers, which they could then contribute to our shared library.

### 3.1 ML is Easier, but not Easy Enough

While we designed our initial system to minimize setup cost, we nevertheless encountered resistance that limited uptake. Traditional supervised machine learning models work well for classification tasks with a pre-determined coding scheme and a high quality labeled dataset. This approach, however, is often time consuming to instantiate and brittle to future changes. For instance, if the input data format shifts or new themes emerge, training data needs to be relabeled and classifiers retrained.

[2]https://www.tensorflow.org/tfx

In our case, the labor cost of manually labeling a training dataset coupled with the technical cost of implementing machine learning classifiers was simply too high. Creating a large enough training dataset for each category required an initial investment of time similar or greater than simply hand-coding the responses, and in many cases under-resourced research teams simply weren't able to make this time investment. Additionally, even in cases where a training set existed, a researcher or engineer with sufficient technical knowledge would have to train, validate, and optimize each classifier, and later implement the pipeline as a whole. We found over the course of our first year that few teams were able to make this investment, despite expressing a clear need for automation.

## 4 Productionized LLM Pipelines

In parallel to our work, we noticed a proliferation of tools that leveraged large language models (LLMs), which were designed to extract themes and apply labels to open text data collected from individual, one-off surveys[6]. These tools either required technical knowledge and resources to run, or they did not provide a streamlined interface for researchers to classify large amounts of tabular data on a continuous basis. While they therefore provided inspiration and direction for our project, they could not fully automate classification from in-product surveys.

In response to the initial barriers which prevented teams from implementing our initial solution, we started to investigate LLMs as a potential alternative approach. LLMs mitigate some of the initial challenges we faced in two key ways. LLMs can both 1) infer themes from open text data without a pre-determined coding scheme, and 2) classify open text data without a labeled training dataset[6]. This effectively means that researchers do not have to create labeled training data to set up a classification pipeline or to apply different labels as new themes emerge.

We therefore designed a second iteration of fully automated classification pipelines that leverage an untuned LLM (Gemini) to automatically classify open text data according to a predefined category scheme. Researchers simply provide a category name and an English language description for each category. The categories are then supplied to the LLM along with boilerplate prompting, and the LLM classifies each response according to the predefined schema.

This implementation has two distinct advantages over our initial, traditional ML approach. First and perhaps most obviously, an untuned LLM does not require labelled data to train, which dramatically reduces the initial time investment required to set up a classifier. Adding or modifying a classification schema only requires adding categories or updating descriptions, whereas traditional machine learning would necessitate an entirely new training dataset. Second, the ability to write prompts thereby "tuning" the classifier using plain English facilitates much tighter collaboration between technical practitioners (in our case quantitative UX researchers), and non-technical stakeholders (e.g. qualitative researchers, other UX practitioners and managers, project managers, among other roles). Whereas creating, tuning, and otherwise modifying our traditional ML pipelines required interacting with our relatively complex auto-ML platform, LLM prompts can be modified by anyone with the domain knowledge to write a succinct category description in English. This not only distributes setup and maintenance burden over a wider group of individuals, but it also allows UX researchers who have expert domain knowledge but limited machine learning experience to participate in the development process.

### 4.1 Performance Trade-offs

While LLMs substantially lower the time and knowledge cost of designing and deploying a classifier, in the specific context of productionized data pipelines these systems have notable drawbacks. Overall, untuned LLMs tend to predict categories less accurately than a traditional machine learning model, although differences in performance are inconsistent[7]. Furthermore, LLM predictions are functionally non-deterministic, meaning that for the same text input they will not always generate the same prediction. This adds to the overall unpredictability and variability in LLM output, which is undesirable for fully automated systems.

In our experience, research teams tend to be hesitant to adopt automated systems without understanding accuracy trade-offs. Researchers who have previously coded open text comments by hand often question the accuracy of the model, and whether it will provide a suitable replacement for a human with expert domain knowledge. While this line of questioning is understandable, to our knowledge the only reliable way to provide performance metrics for a given pipeline is to create a ground-truth test dataset against which to evaluate standard ML performance or inter-coder agreement metrics. Creating these test datasets somewhat diminishes the time-cost advantage of using an LLM over traditional ML.

Nevertheless, we have found that the higher degree of collaboration between researchers facilitated by "tuning" an LLM through plain English prompt engineering often mitigates some of these time-cost barriers. Because researchers can modify prompts in plain English, the entire research team can engage in the process of improving classifications, regardless of an individual's experience with machine learning. This type of engagement helps to build both understanding and trust in the overall system.

## 5 Case Study: Automated Feedback Analysis for a Product Team

In the following section we present our first case study from one product team at Google, which provides a relatively straightforward template for our PoV "play". The process described is relatively context agnostic—i.e. it could be extended to any research team that collects and processes open-text data in order to develop and refine a UX research PoV.

The product team (and many other teams at Google) keep a pulse on user sentiment feedback through the in-product Happiness Tracking intercept Survey. The resulting data is important and is used broadly, both strategically for leadership to understand how our users feel over time, and for the team to improve the product to meet evolving user needs. However, the manual labor required to analyze it is high, and the process can break if there is no researcher or the trained researcher is not available. In the past these situations have sometimes left the team without a consistent source of sentiment feedback.

### 5.1 Process and problems identified

Historically this team has implemented a specific categorization and analysis system that is common across most Google UXR teams, but is not easy to learn quickly. The researcher dedicated to the analysis needs to first learn the common system of processing the data from various documentation sources, learn and remember what the cross organizational common list of categories are and their meanings, regularly categorize large sets of incoming data, then plug the final categorizations into a dashboard for visibility of the most common issues. Finally the researcher identifies specific areas to improve within those issues and works with the team to get commitments to those areas. Not only does the categorization process take time (sometimes multiple days in the ideal case that that is one of the researcher's only priorities), but multiple steps in this process can also introduce errors into the output and/or break the dashboard if not done correctly and consistently every time. If a researcher has to shift priorities, go on leave, etc., the data goes unused and the team cannot use the output for product improvement.

### 5.2 Tool and Process Selection

Given the recent increase in the use of AI tools for research, the UXR responsible for categorization and analysis chose to explore several tools that could help the team decrease the brittleness of the process and increase researcher efficiency. The UXR explored various internal tools, but none of them solved the team's most important needs. One tool came close, but it required too much set up, which would have to be repeated every time analysis was done (such as monthly).

In contrast, the data pipelining project outlined in this paper enabled the product team to set up a prompt and fully automate the process of ongoing categorization and dashboard output. It was a clear choice to move forward with this more streamlined and automated process, as it freed up UXR time and resources while enabling the team to access continuously updated user feedback.

### 5.3 Setup and Iterative Development

Initial system setup was relatively straightforward. The product UXR provided a list of categories and descriptions associated with each category. An example response/description pair might read as follows:

> **Poor performance:** Unstable; Slow/speed; Features/Capabilities working incorrectly or not working at all; Buggy; Less robust; Cold start time; Scaling limitations; Size/Resource intensive; Competitor's performance is better.

Open-text user feedback for this product was already stored in a table, which was programmatically updated on a daily basis. We created an additional daily process that queried this existing table for new data, prompted an LLM to classify each row according to the provided categories, and finally write the output to a new table. This new output table provides the backend for a dashboard that displays code volume over time.

As with every classification process, we observe some disagreement between human and machine classifications. To obtain baseline performance metrics for our classification pipeline, we ingested old responses that had been hand coded by the UXR during the prior manual process and calculated standard machine learning metrics (precision, recall, accuracy, etc). We then iteratively analyzed disagreements between machine and human coders and subsequently tweaked our LLM prompt.

Compared to a traditional ML approach, the vast majority of the "development" effort for this system has been devoted to this iterative process of refining categories, not writing code. This means that, while technical practitioners were involved in the initial setup of the system, UXRs and other stakeholders have substantially more ownership and control over the classificaiton and insight generation process.

## 6 Current and Future Work

Throughout the process of developing automated classification pipelines, it has become clear that we need to streamline the deployment process in two key areas. First, while we reduced the technical barriers associated with building a classifier, deploying these systems still requires a small amount of boilerplate code and configuration. "Vibe coding" this boilerplate with an LLM is one possible solution, but given that this process is relatively static it should also be possible to develop a UI that fully abstracts the deployment process. One such interface is currently under development, which would allow non-technical stakeholders to develop and deploy their own pipelines without writing code.

Second, understanding classifier performance beyond spot checking labels requires some level of machine learning knowledge. While understanding the validity of an instrument or method is important to any researcher, familiarity with machine learning validation metrics (or alternatively inter-rater reliability) varies. Our case initial case study demonstrated that in order to make the category development process more iterative and fluid, researchers need non-technical tooling that also helps them understand the performance of their classification pipelines on test data.

## 7 Conclusion: The Automated Open-text Classification Playcard

By automating open text feedback classification, both UX researchers and a wide range of product stakeholders (project managers, engineers, and directors) are able to leverage rich open text data in order to make decisions about future product development. At Google, these data have previously been inaccessible to most teams due to the time cost required for manual coding and analysis, but automation provides new opportunities to engage with user feedback. Additionally, the increased availability of LLMs has reduced technical investment required to implement automated systems, further reducing barriers preventing the use of open text feedback.

In the PoV framework, automated open-text classification pipelines allow a researcher to effectively move from data collection to the insights generation phase, reducing the time-cost associated with this transition and ultimately helping the researcher to develop a PoV. Because these pipelines are designed to process data on an ongoing basis, they also facilitate iterative updating and refining of existing PoVs as a product and user-base matures. Notably, automated open-text classification does not *replace* the insights generation phase of the process, but rather allows a researcher to make sense of large quantities of data on a continuous, ongoing basis. By lowering the barriers associated with moving between phases, a researcher is more likely to use these data to inform and refine a nuanced PoV.

Automation does come with drawbacks. Researchers often use the coding process to understand their data, and automated systems inherently aim to reduce or eliminate the necessity for this sort of immersion. If automated open text classification systems become more widespread, researchers will need to purposefully engage with their data on a regular basis, both for quality checks and to better understand their users' feedback. Nevertheless, given that many teams are unable to use open text data without automation due to the time cost of hand coding, these systems provide a general step in the right direction

## References

[1] Tom Brown, Benjamin Mann, Nick Ryder, Melanie Subbiah, Jared D Kaplan, Prafulla Dhariwal, Arvind Neelakantan, Pranav Shyam, Girish Sastry, Amanda Askell, Sandhini Agarwal, Ariel Herbert-Voss, Gretchen Krueger, Tom Henighan, Rewon Child, Aditya Ramesh, Daniel Ziegler, Jeffrey Wu, Clemens Winter, Chris Hesse, Mark Chen, Eric Sigler, Mateusz Litwin, Scott Gray, Benjamin Chess, Jack Clark, Christopher Berner, Sam McCandlish, Alec Radford, Ilya Sutskever, and Dario Amodei. 2020. Language Models are Few-Shot Learners. *Advances in Neural Information Processing Systems* 33 (2020), 1877–1901.

[2] Youngjin Chae and Thomas Davidson. 2023. Large language models for text classification: From zero-shot learning to instruction-tuning. *SocArXiv* (Aug. 2023).

[3] Nicolas Antonio Cloutier and Nathalie Japkowicz. 2023. Fine-tuned generative LLM oversampling can improve performance over traditional techniques on multiclass imbalanced text classification. In *2023 IEEE International Conference on Big Data (BigData)*. IEEE, 5181–5186.

[4] Huseyin Dogan, Stephen Giff, and Renee Barsoum. 2024. User experience research: Point of View playbook. In *Extended Abstracts of the CHI Conference on Human Factors in Computing Systems*, Vol. 1. ACM, New York, NY, USA, 1–7.

[5] Paul Trust and Rosane Minghim. 2024. A study on text classification in the age of large language models. *Mach. Learn. Knowl. Extr.* 6, 4 (Nov. 2024), 2688–2721.

[6] Alec Radford, Jeff Wu, R Child, D Luan, Dario Amodei, and I Sutskever. 2019. Language Models are Unsupervised Multitask Learners. (2019).

[7] Zhiqiang Wang, Yiran Pang, and Yanbin Lin. 2023. Large language models are zero-shot text classifiers. *arXiv [cs.CL]* (Dec. 2023).